\documentstyle[11pt,aasms4,epsf]{article}

\def\folio{\ifnum\pageno<2\nopagenumbers\else\number\pageno\fi}
\newtoks\headline \headline={\hss\twelverm\folio\hss} % page no at top of page
\newtoks\footline \footline={{\hfil}} % no footer
\font\mathbf=cmmib10 scaled 1000             % bold math font for tenpoint
\def\ref{\par\noindent\hangindent=2pc \hangafter=1 }
\def\amin{\ifmmode^{\prime}\else$^{\prime}$\fi}
\def\asec{\ifmmode^{\prime\prime}\else$^{\prime\prime}$\fi}

\def\etal{{et al. }}
\def\cappage #1 #2 #3 {\vfill\eject\pageno=#1
\vglue 10 true in minus 10 true in \noindent{\bf Figure #2.} #3}
\def\ee #1 {\times 10^{#1}}
\def\ut #1 #2 { \, \hbox{#1}^{#2}}
\def\u #1 { \, \hbox{#1}}

\def\kms {\, \hbox{km}\,\hbox{s}^{-1}}

\let\grad=\nabla
\def\cross{{\bf \times}}
\def\curl #1 {\grad \cross #1}
\def\div #1 {\grad \cdot #1}

\def\kms    {\hbox{km{\hskip0.1em}s$^{-1}$}}    % km/s
\def\etal   {{\it et al. }}                     % et al.

\begin{document}

\title{Thermal OH (1667/65 MHz) Absorption  and Nonthermal OH (1720 MHz)
Emission Towards the  W28 Supernova Remnant}

\author{F. Yusef-Zadeh}
\affil{Department of Physics and Astronomy, Northwestern University,
Evanston, IL. 60208 (zadeh@northwestern.edu)}

\author{M. Wardle}
\affil{Department of Physics, Macquarie University, NSW 2109, Australia
(wardle@physics.mq.edu.au)}

\author{D.A. Roberts}
\affil{Department of Physics and Astronomy, Northwestern University,
Evanston, IL. 60208 \& Adler Planetarium, 1300 S. Lake Shore Drive,
Chicago, IL. 60605 (doug-roberts@northwestern.edu)}

\begin{abstract}

The W28 supernova remnant is an  excellent prototype for observing
 shocked gas resulting from the interaction of supernova remnants
(SNRs) and  adjacent molecular clouds (MCs).  We present two new
signatures of shocked molecular gas in this remnant.  One is the
detection of main-line extended OH (1667 MHz) absorption with broad
linewidths.  The column density of OH estimated from the optical depth
profiles is consistent with a theoretical model in which 
 OH  is formed behind a C-type shock front.  The second
is the detection of extended, weak OH (1720 MHz) line emission with
narrow linewidth distributed throughout the shocked region of W28.
These give observational support to the idea that compact maser
sources delineate the brightest component of a much larger region of
main line OH absorption and nonthermal OH (1720 MHz) emission tracing
the global structure of shocked molecular gas.  Main line OH (1665/67)
absorption and extended OH (1720 MHz) emission line studies can serve
as powerful tools to detect SNR-MC interaction even when bright OH (1720
MHz) masers are absent.

\end{abstract}

\keywords{ISM: Clouds---ISM: general---shock waves---supernova
remnants---X-rays: ISM}

\section{Introduction}

OH (1720 MHz) masers unaccompanied by masing in the 1665 and 1667 MHz
main lines occur where supernova remnants (SNRs) interact with
molecular clouds (MCs) (Frail, Goss \& Slysh 1994; Yusef-Zadeh et al.\
1996; Green et al.\ 1997; Wardle and Yusef-Zadeh 2002).  These
so-called SNR masers are collisionally pumped within a shock wave
driven into a dense molecular cloud.  The OH molecule is believed to
be produced by the indirect dissociation of water formed within a
C-type shock (Lockett, Gauthier \& Elitzur 1999) by X-ray photons from
the interior of the supernova remnant (Wardle 1999).

Despite several radio surveys to find SNR-MC interaction sites in
the Galaxy, OH (1720 MHz) masers have been detected toward only 19 SNRs,
 about 10\% of the total
number of observed SNRs (Frail et al.  1997; Koralesky et al.  1998;
Green et al.  1997; Yusef-Zadeh et al.  2002). 
 This low detection
rate is partly due to the fact that OH (1720 MHz) masers only arise under
restrictive conditions in density and temperature (Lockett et al.
1999) and that velocity coherence and maximum pathlength for
amplification are achieved at the limb of the remnant where the shock
propagates in a direction perpendicular to the line of sight.  However, the 
OH
abundance increases throughout the shocked region and is expected to
be more extensive and less restrictive in its physical conditions
(Wardle 1999) than the regions where bright maser spots occur.

The OH column densities required for the formation of OH (1720 MHz)
masers ($\ga 10^{16}\ut cm -2 $) should be sufficient to create detectable
thermal OH absorption in the main line transitions of OH (1665/67 MHz)
molecule if the background continuum is strong.  In order to test this
idea, we observed a  nearby interacting SNR, W28 at
the distance of 1.8 kpc (Rho et al.  1994).  W28 is a  prototype SNR 
maser with
numerous compact and bright maser spots distributed throughout the
remnant (Claussen et al.  1997).  OH absorption towards W28 has
previously been detected with single-dish observations (Pastchenko \&
Slysh 1974) and with the Australian Telescope Compact Array (Green et
al.  2000).  Here we report the VLA detection of broad OH (1667 MHz)
absorption lines against the prominent W28 SNR  continuum
source.  The column density of shocked OH gas estimated from optical
depth profiles toward W28 is consistent with the theoretical
prediction.  In addition, we present extended, weak OH (1720 MHz)
maser emission from W28.  We then suggest that studies of thermal OH (1667 
MHz)
absorption and extended nonthermal OH (1720 MHz) emission  can be used to 
search
for additional SNR-MC interactions where compact 1720 MHz masers are
absent.

\section{Observations}

The Very Large Array (VLA) of the National Radio Astronomy
Observatory\footnote{The National Radio Astronomy Observatory is a
facility of the National Science Foundation, operated under a
cooperative agreement by Associated Universities, Inc.} was used on
October 7, 2001 and July 18, 2000 to observe  OH (1667 MHz) absorption and 
OH(1720 MHz) emission line studies, respectively.  We used the compact
D  configuration to observe OH (1667 MHz) line absorption toward
W28 at center velocities of 10 \kms.  
 Mode 4 of
the correlator was  used with 64 channels per IF; each IF pair is tuned
to the different line transitions of OH, (1667 MHz and 1720 MHz).  The
channel separation was 6.104 kHz, which corresponds to 0.55 km/s.
Standard calibrations of complex gains and bandpass were  carried out
using 1328+307, 1730--130 and 1851--235. 
 Because of a strong negative bowl produced surrounding
the continuum image of W28, a zero-spacing flux of 60 Jy was added to
the continuum {\it uv} data.  The {\it uv} data were then
self-calibrated before the final line (l) and continuum (c) images
were CLEANed.  The rms noise in a given channel of W28 is  5.5
mJy beam$^{-1}$.  An optical depth ($\tau$) image and
corresponding error images were constructed for W28 by using $\tau =- ln
((l + c)/c)$.  

In the second VLA observation, we carried out an OH (1720 MHz)
emission line study of W28.  Unlike the previous high-resolution
snapshot OH (1720 MHz) observations of W28 using the A-configuration
of the VLA (Frail et al.  1994; Claussen et al 1997), we searched for
extended, weak OH (1720 MHz) emission from W28 using the  compact 
D configuration with a relatively uniform {\it uv} coverage so as to be
sensitive to extended but narrow line OH (1720 MHz) emission features.
Mode 2AC of the correlator was chosen with 128 channels, a channel
width of 1.06 \kms\ and a total bandwidth of 0.7812 MHz.  This
observation used similar phase and flux calibrators to that of
absorption line observations.  The rms noise for a given channel is
6.2 mJy beam$^{-1}$.  The rms noise for three velocity channels at 9.6, 10.7 and
11.8 \kms\ where the most intense maser emission is observed were three
times higher.

\section{Results}

\subsection{W28 (SNR G6.4-0.1)}
\subsubsection{Broad Thermal OH (1667 MHz)  Absorption}

The right panel of Figure 1 shows a contour representation of OH (1667 
MHz)
line absorption toward W28 integrated between --3.2 and 23.7 \kms.
The grayscale image shows the continuum image of W28 constructed 
from line-free channels.  Much of the
absorption is distributed prominently to the northeast of the remnant
where background continuum is strongest.  The absorbing gas velocities
are very similar to OH (1720 MHz) maser velocities ranging between +4.8
and +16 \kms\ as reported by Claussen et al.  (1997) and Frail et al.
(1994).  The panel to the left of the remnant in Figure 1 shows a
typical absorption spectrum (position is indicated by the cross on 
the right panel) 
with a total linewidth  of $\approx$30 \kms.

Figure 2 shows a grayscale continuum image at 20cm (Dubner et al.
2000) with the distribution of compact OH (1720 MHz) maser
spots represented as crosses (Claussen et al.  1997) superposed.  The insets to
the left and top of Figure 2 show the fitted optical depth profiles
and their corresponding residuals.  The optical depth velocity
profiles coincide with clusters of OH (1720 MHz) maser sources
identified by Claussen et al.  (1997).  Gaussian fitted spectra of the
top and left $\tau$ profiles show center velocities between V$_{LSR}$
= 10.2 and 8.1 \kms,  which are similar to the velocities of of OH (1720
MHz) maser spots identified in the F, C and D regions of Claussen et
al.  (1997).  The linewidths (FWHM) of 14 and 11 \kms\ are much larger
than typical linewidths of ambient molecular clouds, thus confirming
the interaction picture of W28 with its surrounding molecular cloud
(e.g. Wootten 1981).  Table 1 gives the parameters of the fits to 
four
$\tau$ spectra, all of which show broad line widths ranging between 7
and 10 \kms.  However, the top-right spectrum of Figure 2 shows two
narrow absorption lines with $\tau$=0.23 and 0.28 plus a broad
absorption feature with $\tau$=0.08 at centroid velocity of 10 \kms.
This spectrum is seen against a bright compact continuum source
G6.6-0.1 $\alpha, \delta$ (J2000)= $18^{\rm h} 00^{\rm m} 49.8^{\rm s}, -23^\circ 20'
24''.56$   showing a flux density of 0.76 Jy near the center of W28.
This source is known to have a flat spectral index (Andrews et al.
1983; Dubner et al.  2000).  The line profiles show FWHM linewidths of 
1.5--1.8 \kms\ and are centered at 12 and 18 \kms.  The narrow line
widths of absorption features suggest that they are not associated
with the SNR-MC interaction.  The high velocity component is close to
the high velocity component of HI absorption measurements
(Radhakrishnan et al.  1972).  Claussen et al.  (1997) suggest that
the systemic velocity of the molecular cloud associated with W28 is at
17.6 \kms\ and that the low-velocity 7 \kms\ component is being
accelerated by the SNR shock along the line of sight by about --10
\kms. This suggests that the narrow line spectra observed in Figure
2 are associated with the MC in the vicinity of W28. 

A bright,  compact OH (1667 MHz) maser source with two velocity
components is detected in the vicinity of W28.  This source 
lies at
$\alpha, \delta$ (J2000)= $18^{\rm h} 01^{\rm m} 57.78^{\rm s}, -23^\circ 12' 31''.43$.
The spectrum shows two narrow velocity components at 17.1 \kms\ and
23.5 \kms\ with a flux density of 0.39 and 0.05 Jy, respectively,
resembling the spectrum of OH/IR stars.

\subsubsection{Extended Maser Emission}
Figure 3
shows contours of the OH(1720 MHz) line emission integrated
between 5.3 and 21.3 km/s (with the exception of three channels at
9.6, 10.6 and 11.7 \kms, which were contaminated by saturation from
strong maser emission) superposed on a grayscale continuum image
of W28.
 The crosses indicate the positions of 41 compact maser sources
listed by Claussen et al.  (1997).  We note extended and compact OH
(1720 MHz) emission from W28.  The top two insets show 
the velocity profiles
of two isolated sources to the north of the shell.  
These sources, as
well as extended maser emission adjacent to compact maser sources,  have
not been detected in earlier high resolution observations which over-resolved 
these  extended features. 
  The third 
OH (1720 MHz) velocity profile in 
the bottom left panel of Figure 3 coincides with the position of the 
optical depth profile presented in the top left panel of Figure 2. It is 
clear that where OH(1667 MHz) absorption profile is broad, OH (1720 MHz) 
line emission is narrow.

 The extended OH (1720 MHz) emission features
have typical linewidths of about 1-3 \kms.  The flux density of
extended features are typically about 100-300 mJy/beam corresponding
to a brightness temperature T$_b\sim$ 20--60 K. The kinematics of
extended OH (1720 MHz) emission is very similar to that of compact
masers,   suggesting that they 
co-exist with each other. The extended features with  
their  nonthermal characteristics trace shocked CO
(3--2) molecular gas toward W28 (Arikawa et al.  1999) in spite of
their low brightness temperatures.

\section{Discussion}
The column density of OH toward W28 may be estimated from the observed
optical depth velocity profiles using
\begin{equation}
	N_{\mathrm{OH}} = 2.2785 \times 10^{14}\ T_{\mathrm{ex}} \int \tau_v\
	dv \ \ \ut cm -2
\end{equation}
for the 1667 MHz line once a value for the excitation temperature,
$T_{\mathrm{ex}}$, has been chosen (Crutcher 1977).  Because the 
energy spacing
between the upper and lower levels is $\approx$0.08 K,
$T_{\mathrm{ex}}$ is sensitive to small departures of the level
populations from LTE. In order to estimate $T_{\mathrm{ex}}$, we
modelled OH level populations in a uniform slab of gas with a FWHM
linewidth of 10 \kms.  $T_{\mathrm{ex}}$ is determined largely
by the kinetic temperature $T_k$ and is insensitive to density and
column density for the parameters that are relevant here.  The
excitation temperature is close to the kinetic temperature $T_k$ for
$T_k\la 20\u K $ and then decreases with increasing kinetic temperature as
collisions begin to populate the higher rotational states:
$T_{\mathrm{ex}}=18$, 8, and 3 K at $T_k =30$, 50 K and 75 K
respectively.  Therefore we adopt $T_{ex}=10$\,K as a nominal value
with the assumption that 
$T_{ex}$  may differ by a factor of two in either 
direction.
Our measurements indicate that $\int \tau_v\ dv \approx 10$ \kms,
therefore $N_{\mathrm{OH}}\approx 2\ee 16 \ut cm -2 $.  This OH column
density is consistent with that required to produce 1720 MHz maser
emission (Lockett, Gauthier \& Elitzur 1999) and expected from X-ray
dissociation of water produced in a C-type shock wave driven into the
molecular cloud by the supernova remnant (Wardle 1999).

Extended 1720 MHz emission can be produced by weak amplification in
the OH behind a \textit{face-on} shock front as opposed to the edge-on
geometry required by the bright maser spots.  The brightness
temperature of the 1720 MHz emission, $\sim 30 \u K $ at line 
center,
is roughly equal to that of the background continuum.  The implied
amplification factor of 2 requires an OH column
$N_{\mathrm{OH}}\approx 10^{15} \ut cm -2 $ with $T_k \ga 50 \u K $
(Lockett et al 1999), and line FWHM $\la 0.5$\kms, conditions that are
achieved in the cooling tail of a C-type shock wave (Wardle 1999).  It
should be possible to detect this weakly-masing component in
absorption at 1667 MHz, as the optical depth is $\approx 2$ at line
center. H$_2$CO absorption study of  W28  also indicates that the 
interacting cloud has a  low kinetic temperature and that the properties of 
the interacting cloud are  consistent with 
the picture of cold post-shocked  gas (Slysh et al.  1980).

The observations presented here trace extended regions of shocked molecular
gas in which the abundance of OH has been enhanced.  Our results suggest
two new ways to search for SNR-MC interaction sites where OH (1720 MHz)
masers are absent.  One is to search for extended OH (1720 MHz) maser
emission produced by shock waves.  Unlike bright, compact masers this
emission does not require the shock to propagate almost perpendicular to
the line of sight, and should therefore be more generic.  This emission has
been detected in the five SNRs with compact masers that have been searched
to date (Yusef-Zadeh \etal 1995; 1999; and this paper).  SNR--molecular
cloud interaction sites can also be identified by searching for broad OH
absorption lines at 1665/67 MHz delineating the region where shocked
molecular gas overlies the radio continuum emission from the SNR. A strong
empirical association between X-ray emitting centrally-filled
mixed-morphology remnants and maser-emitting shell-type SNRs has recently
been demonstrated by Yusef-Zadeh et al.  (2002).  Remnants that show
centrally filled thermal X-ray emission are strong candidates for such
studies.

Acknowledgments: We thank G. Dubner for providing us
with the  continuum image of W28.

\vfill\eject

\begin{figure}
%\plotone{fig1_reduced.ps}
\figcaption{ [right panel] Contours of OH (1667 MHz) absorption toward W28
integrated between $-2.1$ and 23.7 \kms\ at $-4.0$, $-2.0$, $-1.0$, $-0.5$ 
 mJy  beam$^{-1}$  \kms\
are  superposed on a grayscale continuum image with a
spatial resolution
of 68$''\times30''$, PA=40$^\circ$. The range is grayscale brightness  is
between 0 and 270  mJy.
[left panel] A typical spectrum showing an OH (1667 MHz) absorption line
toward
the cross at 
$\alpha, 
\delta$ (J2000)= $18^{\rm h} 01^{\rm m} 41.43^{\rm s}, -23^\circ 25' 11''.3$.}

\end{figure}

\begin{figure}
%\plotone{fig2.ps}
\figcaption{A  grayscale continuum image at 20cm with a
spatial resolution
of
88$''\times48''$ (Dubner et al. 2000) with
the distribution of compact OH (1720 MHz) maser spots
represented as crosses  (Claussen et al. 1997) superposed.
The insets show the derived optical depths (including error bars),
Gaussian fits, and corresponding residuals.}
\end{figure}

\begin{figure}
%\plotone{fig3.ps}
\figcaption{ A grayscale continuum image of W28 at 20cm (Dubner et al. 
2000) with superposed 
contours of OH (1720 MHz) maser emission set at $-3$, 3, 5, 7, 9, 11, 13, 15, 20, 25, 
30, 40, 50, 60, 70, 80, 100 mJy beam$^{-1}$ with a spatial resolution of 
86.6$''\times34.6''$, PA=14$^\circ$ (shown as the white ellipse in the
bottom right hand corner).  The 
crosses represent the
position of compact OH (1720 MHz) masers identified by Claussen et al.
(1997). The respective positions of the three spectra on the top, bottom 
left, 
and top left are: 
$\alpha, \delta$ (J2000)= $18^{\rm h} 01^{\rm m} 43.6^{\rm s}, -23^\circ 24' 19''.8$, 
$18^{\rm h} 01^{\rm m} 43.33^{\rm s}, -23^\circ 12' 59''.2$ and 
$18^{\rm h} 01^{\rm m} 32.4^{\rm s}, -23^\circ 11' 10''.01$.
}
\end{figure}

\vfill\eject

%\documentstyle[apjpt4]{article}

%\begin{document}

%
% My macros
%
\def\shsp   {\hskip0.1em}                       % A bit bigger, but small
\def\x      {$\times$}                          % Times symbol
\def\deg    {$^{\circ}$}                        % Degrees symbol
\def\amin   {$^{\prime}$}                       % Arminutes symbol
\def\asec   {$^{\prime\prime}$}                 % Arcseconds symbol
\def\hour   {$^{\rm h}$}                        % Hours of time
\def\min    {$^{\rm m}$}                        % Minutes of time
\def\damin  {\hbox{$.\mkern-4mu^\prime$}}       % Arcminutes over dot
\def\dasec  {\hbox{$.\!\!^{\prime\prime}$}}     % Arcseconds over dot
\def\dsec   {\hbox{$.\!\!^{\rm s}$}}            % Second over dot
\def\vlsr   {\hbox{${V_{\rm LSR}}$}}            % Vlsr
\def\thco   {\hbox{$^{13}$CO}}                  % 13CO
\def\ceto   {\hbox{C$^{18}$O}}                  % C18O
\def\kms    {\hbox{km{\hskip0.1em}s$^{-1}$}}    % km/s
\def\perbeam {\hbox{beam$^{-1}$}}               % /beam

\vfill\eject

\begin{deluxetable}{lllll}
\tablenum{1}
\tablecolumns{5}
\tablewidth{0pt}
\tablecaption{OH (1667 MHz) Optical Depth towards W28}
\tablehead{
\colhead{Position} &
\colhead{Amplitude} &
\colhead{Center Velocity} &
\colhead{FWHM Line Width} &
\colhead{Corresponding OH} \nl
\colhead{(RA, Dec (J2000))} &
\colhead{(optical depth)} &
\colhead{(km s$^{-1}$)} &
\colhead{(km s$^{-1}$)} &
\colhead{(1720 MHz) masers} \nl
}
\startdata
18\hour00\min49\dsec924, --23\deg20\amin26\dasec66 &
0.229 $\pm$ 0.017 &
12.471 $\pm$ 0.130 &
4.439 $\pm$ 0.387 &
A \nl
&
0.276 $\pm$ 0.018 &
18.870 $\pm$ 0.093 &
3.328 $\pm$ 0.253 &
\nl
&
0.085 $\pm$ 0.011 &
9.983 $\pm$ 1.156 &
23.980 $\pm$ 2.687 &
\nl
18\hour01\min50\dsec002, --23\deg18\amin46\dasec58 &
0.726 $\pm$ 0.027 &
10.269 $\pm$ 0.183 &
14.163 $\pm$ 0.765 &
F\nl
18\hour01\min44\dsec041, --23\deg24\amin23\dasec05 &
1.148 $\pm$ 0.075 &
6.845 $\pm$ 0.195 &
8.928 $\pm$ 0.489 &
C, D\nl
18\hour01\min41\dsec444, --23\deg25\amin05\dasec26 &
0.668 $\pm$ 0.026 &
8.533 $\pm$ 0.170 &
11.508 $\pm$ 0.532 &
C, D\nl
\enddata
\end{deluxetable}

%\end{document}
%\vfill\eject

\end{document}